\documentclass[11pt,a4wide]{article}
\usepackage{newlfont}
\usepackage{amsmath}
\usepackage{graphics}
\usepackage{float}
\usepackage{graphicx}
\usepackage{epsfig}
\usepackage{multirow}

\begin{document}

\title{Conventional and Hybrid $B_c$ Mesons in an Extended Potential Model}
\author{Nosheen Akbar\thanks{e mail: nosheenakbar@cuilahore.edu.pk,noshinakbar@yahoo.com}, M. Atif Sultan \thanks{e mail: atifsultan.chep@pu.edu.pk}, Bilal Masud\thanks{e mail: bilalmasud.chep@pu.edu.pk}, Faisal Akram \thanks{e mail: faisal.chep@pu.edu.pk} \\
\textit{$\ast$COMSATS Institute of Information and Technology,
Lahore(54000), Pakistan.} \\
\textit{Centre For High Energy Physics, University of the Punjab,
Lahore(54590), Pakistan.}}
\date{}
\maketitle

\begin{abstract}
Using our analytical expressions that well model the lattice simulations of the gluonic excitations, we use the extended quark potential model to study the effects of orbital and radial excitations on the masses and sizes of conventional and hybrid $B_c$ mesons. A non relativistic formalism is used to
 numerically calculate the wave functions using the shooting method; this allows us also calculating the  $E1$, $M1$ radiative partial widths for  conventional meson to meson and hybrid to hybrid transitions. We incorporate spin mixing and compare our calculated spectrum and decay widths with the available experimental $B_c$ masses and the theoretically predicted spectra and the decay widths by other groups. Our results can help consider both conventional and hybrid quantum numbers to $B_c$ mesons as experimental results become available.

 \end{abstract}


\section*{I. Introduction}

Once we have written potential for a two body system, we can use this to solve a relativistic or even non-relativistic wave equation and
then use the resulting wave functions and energies to predict properties of the system. In contrast to the electromagnetic field,
 the total energy of the color (or the gluonic) field for a set of positions of a quark and an antiquark may have more than one value; computer simulations of quantum chromodynamics (QCD) produce a number of curves for the total energy of the gluonic field~\cite{morningstar99}. A possible solution to the resulting difficulty is to keep defining potential energy as a function of the quark and antiquark positions, but use different potentials for each variety of the gluonic field for one set of quark and antiquark positions.
For the gluonic field in its ground state denoted by $\Sigma^+_g$ in \cite{morningstar99}, a number of expressions are known including the Cornell potential \cite{Eichten 1980} of the Coulombic plus linear form.
To this spin dependent terms, like the spin-spin interaction~\cite{charmonia05}, can be added. If this potential is used in a non-relativistic Schr$\ddot{\text{o}}$dinger equation,
relativistic effects can be incorporated to a large extent by adjusting values of the constituent quark masses.
For the gluonic field in the first excited state $\Pi_u$, the above potential
can be used along with an additional term to model the difference between the first gluonic
excitation and ground state of the gluonic field. We suggested in ref.\cite{Nosheen11} a number of analytical expressions
for this potential and then used the one which best fits the relevant lattice-generated discrete
energies to find a number of dynamical implications (radii, wave functions at origin, leptonic and two photon decay widths, $E1$ and $M1$ radiative transitions) for heavy quarkonia that can be compared with actual hard experiments. These sectors have zero net flavor and hence are eigenstates of $C$ parity.
Now, we extend this work to a sector of net non zero flavor, namely $B_c$ where $C$ parity is not a good quantum number and hence here
states with different total spins but with the same total angular momentum can mix.
Such mesons cannot annihilate into gluons. So these are more stable; their widths are less than a hundred keV.
Thus our radiative corrections ($E_1$ and $M_1$ transitions) can be compared with experiments without combining with annihilation diagrams.

Experimentally only two $B_c$ meson states ($B_c(1S),B_c(2S)$) are discovered with mass $6.2749 \pm 0.0008$ GeV and $ 6.842 \pm 0.004 \pm 0.005$ GeV respectively.
 Many phenomenologists are working to investigate the nature of $B_{c}$ mesons. Spectrum of $B_c$ meson is calculated by using the quark potential model~\cite{9402210,9511267,9806444,0210381,0406228,1504.07538}, the heavy quark effective theory~\cite{9412269}, QCD Sum rule~\cite{9406339, 1306.3486}, QCD spectral sum rules~\cite{9403208}, and lattice QCD~\cite{0409090,Lattice,0305018}. Refs.~\cite{9402210,9406339, 0406228} compute electromagnetic and hadronic transition rates of $B_c$ mesons, and refs.~\cite{9511267,9806444, 0210381} give predictions for their electromagnetic transition widths. Ref.~\cite{9403208} discusses decay constants and semileptonic widths of mesons with charm and beauty quarks; ref.~\cite{0406228} also reports semileptonic widths of the $B_c^+$ meson.
Both spectrum and decays are used to try identifying a meson. A possibility is that the meson under study is a hybrid. For the $B_c$ sector, hybrids are considered so far only in ref.~\cite{1306.3486}. This work reports, in addition to the spectrum, decay pattern for six states ($0^+, 0^-, 1^+, 1^-,2^+,2^-$). But we predict a more comprehensive list of masses, radii and radiative transitions of hybrid $B_c$ states and pave the way for considering the hybrid option in future studies of mesons with non-zero net flavor as well.

The paper is organized as follows. In the section II, the Schrodinger equation along with the potential models for conventional and hybrid mesons is written. The expressions used to find masses, root mean square radii and $M1$ and $E1$ radiative transition widths for conventional and hybrid $B_c$ mesons are written in section III.
Results for the masses and root mean square radii for the radial and orbital ground and excited states of conventional and hybrid $B_c$ mesons are reported in section IV. Radiative partial widths are also reported in this section.
\qquad

\section*{II. Schrodinger Equation for Conventional and Hybrid $B_c$ mesons}
To calculate the wave function of the bound state of quark-antiquark
pair, we use the radial Schr$\ddot{\text{o}}$dinger equation
\begin{equation}
U^{\prime \prime }(r)+2\mu (E-V(r)-\frac{<L^2_{q \overline{q}}>}{2\mu r^{2}})U(r)=0, \label{P24}
\end{equation}
where $E$ is the energy of meson, $U(r)=rR(r)$ in which $R(r)$ is
the radial factor of the wave function, and $<L^2_{q \overline{q}}>$
is quark-antiquark relative angular momentum given as
\cite{morningstar99,Juge99}
\begin{equation}
\left\langle L_{q\overline{q}}^{2}\right\rangle =L(L+1)-2\Lambda
^{2}+\left\langle J_{g}^{2}\right\rangle.
\end{equation}
For conventional mesons $ \left\langle
L_{q\overline{q}}^{2}\right\rangle =L(L+1)$ with $-2\Lambda
^{2}+\left\langle J_{g}^{2}\right\rangle = 0$ \cite{morningstar99}.
$V(r)$ is the potential defined below.

\subsection*{IIa. Conventional meson Potential}

For the conventional heavy-light mesons, we use the following
potential

\small
\begin{multline}
V(r) = V_{q \overline{q}}(r)= \frac{-4\alpha _{s}}{3r} + b r + \frac{32\pi \alpha_s}{9 m_q m_{\overline{q}}} (\frac{\sigma}{\sqrt{\pi}})^3 e^{-\sigma ^{2}r^{2}} \textbf{S}_{q}. \textbf{S}_{\overline{q}}+\frac{4 \alpha _{s}}{m_q m_{\overline{q}} r^3} T  \\ +(\frac{\textbf{S}_{q}}{4 m_q^2}  + \frac{\textbf{S}_{\overline{q}}}{4 m_{\overline{q}}^2}).\textbf{L} (\frac{4\alpha _{s}}{3r^3}- \frac{b}{r})+ \frac{\textbf{S}_{q}+ \textbf{S}_{\overline{q}}}{2 m_q m_{\overline{q}}}.{\textbf{L}}\frac{4\alpha _{s}}{3r^3}, \label{Vr}
\end{multline}
\normalsize where $\alpha _{s}$ and  $b$ are the strong coupling constant
and string tension respectively, and $T$ is the tensor operator defined
as
\begin{equation}
T=\textbf{S}_q.{\hat{r}}\textbf{S}_{\overline{q}}.{\hat{r}}-\frac{1}{3}\textbf{S}_{q}.
\textbf{S}_{\overline{q}},
\end{equation}
such that
\begin{equation}
<^{3}L_{J}\mid T\mid ^{3}L_{J}>=\Bigg \{
\begin{array}{c}
-\frac{1}{6(2L+3)},J=L+1 \\
+\frac{1}{6},J=L \\
-\frac{L+1}{6(2L-1)},J=L-1.
\end{array}
\end{equation}
Here $L$ is the relative orbital angular momentum of the quark-antiquark
and $S$ is the total spin angular momentum. The spin-orbit potential and
the tensor term are both zero~\cite{charmonia05} for $L=0$, where in the third term $\overrightarrow{S}_{q}.
\overrightarrow{S}_{\overline{q}}=\frac{S(S+1)}{2}-\frac{3}{4}$.
$\mu $ is the reduced mass of the quark and antiquark and $m_{q}$ is
the constituent mass of the quarks.
\subsection*{IIb. Hybrid meson potential}
To describe hybrid meson in the Born-Oppenheimer (BO) approximation
used in \cite{Nosheen11,Juge99,Braaten,Nosheen14}, we use the static
potential $ V_{q\overline{q}}^{h}(r)$ in place of $V(r)$ of eq.
(\ref{Vr}):
\begin{equation}
V_{q\overline{q}}^{h}(r)=V_{q\overline{q}}(r)+V_{g}(r),  \label{hypot}
\end{equation}

\noindent where $V_{g}(r)$ is the gluonic potential whose functional
form varies with the level of gluonic excitation. This potential and
the corresponding gluonic states are labeled by Greek letters
$\Sigma, \Pi, \Delta, ...$ corresponding to $\Lambda =0,1,2... $
which represents the projection of total angular momentum of gluons
onto the quark anti-quark axis. The gluonic states which are even (odd)
under the combined operation of charge conjugation and spatial
inversion are represented by a subscript $g(u)$ to
the label. In
present work we study the hybrids in which the gluons are in the
first excited state, i.e., $\Lambda =1$. This state is represented
by the label $\Pi _{u}$ for which  the squared gluon angular momentum $
\left\langle J_{g}^{2}\right\rangle =2$ and $\Lambda =1$~\cite{Kuti97}
making $-2\Lambda ^{2}+\langle J_{g}^{2}\rangle =0$. For this
the parity of hybrid meson is given by
\begin{equation}
P=\epsilon (-1)^{L+\Lambda +1},
\end{equation}
\noindent where $\epsilon =\pm 1$ for the $\Pi _{u}$ state~\cite{Kuti97}. In the present work we use the following $V_{g}(r)$
\begin{equation}
V_{g}(r)=\frac{c}{r}+A\times e^{-Br^{0.3723}},  \label{vg}
\end{equation}
\noindent where the values of the constants $A=3.4693$ GeV, $B=1.0110$
GeV, and $c=0.1745$ are fixed by our earlier fit~\cite{Nosheen11} to
the lattice data \cite{Kuti97}. It is shown in ref. \cite{Nosheen11}
that the form of eq. (\ref{vg}) provides best fit to the lattice data~\cite{Kuti97}.

\subsection*{IIc. Mixed States}
The mesons with equal quark anti-quark mass satisfy the following parity and charge expressions
 \begin{equation}
P=(-1)^{L+1}\quad \mathrm{and} \quad C=(-1)^{L+S}.
\end{equation}
But mesons with unequal quark anti-quark flavors, like $B_c$ mesons,
are not eigenstates of the charge conjugation. So the states with
different total spins $(S)$ and same total angular momentum $(J)$
can mix. For example, $^1 P_1$ and $^3 P_1$ states of $B_c$ mesons
can mix because both states have same $J=1$, but $S=0$ for $^1 P_1$
and $S=1$ for $^3 P_1$. The measurable $P$ states with $J=1$ are the
linear combinations of $^1 P_1$ and $^3 P_1$ expressed as
\begin{equation}
n P^{'}= n ^1P_1 \mathrm{cos} \theta _{nP} +n ^3P_1 \mathrm{sin} \theta _{nP},
\end{equation}
\begin{equation}
n P = - n ^1P_1 \mathrm{sin} \theta _{nP} +n ^3P_1 \mathrm{cos} \theta _{nP},
\end{equation}
where $\theta_{np}$ is the mixing angle. Similarly experimental $D$
states with $J=2$ are the linear combination of $^1 D_2$ and $^3
D_2$. For the $D$ meson mixed states, linear combinations are
\begin{equation}
n D^{'}= n ^1D_2 \mathrm{cos} \theta _{nD} +n ^3D_2 \mathrm{sin} \theta _{nD},
\end{equation}
\begin{equation}
n D = - n ^1D_2 \mathrm{sin} \theta _{nD} +n ^3D_1 \mathrm{cos} \theta _{nD}.
\end{equation}
For heavy quarks, the mixing angle becomes~\cite{ebert10}
\begin{equation}
\theta_{m_Q \rightarrow \infty}= \tan^{-1}\sqrt{\frac{L}{L+1}},
\end{equation}
so $\theta _{n P} = 35.3^o$ and $\theta_{n D} = 39.2^o$.
\section*{III. Properties of Conventional and Hybrid $B_c$ mesons}
\subsection*{IIIa. Spectrum of Mesons}
To compute the spectrum of mesons, we find numerical solutions of
the Schr$\ddot{\text{o}}$dinger equation by using the shooting
method. The mass of a quark-antiquark meson state is obtained by
addition of constituent quarks mass to the energy $E$ corresponding
to the accepted solutions. The parameters ($\alpha_s, b,\sigma,
m_b$) used in above mentioned conventional meson potential are found
by fitting to the experimentally known $B_c$ mesons ($ B_c(1S) =
6.2749 \pm 0.0008$ GeV and $ B_c(2S) = 6.842 \pm 0.004 \pm 0.005$
GeV).
 We obtain the following values: $\alpha_s=
0.48$, $\sigma=1.0946$ GeV, $b=0.137$ $\text{GeV}^2$. The masses
$m_c= 1.4794$ GeV, $m_b = 4.825$ GeV are taken from Ref.
\cite{Nosheen14,Nosheen16}

\subsection*{IIIb. Radii}
 The normalized wave functions are used to calculate root mean square radii using
the following relation:
\begin{equation}
\sqrt{\langle r^{2}\rangle }=\sqrt{\int U^{\star }r^{2}Udr}.  \label{P25}
\end{equation}
It is noted that terms in the potential which are proportional to
$\frac{1}{r^{3}}$ make the wave function unstable at small distance
whenever $J=L$ or $J=L-1$. In calculating the masses the problem is
overcome by treating these terms through the perturbation theory.
However, calculating the perturbative correction to the wave
function is difficult as in this case the contributions come from
all possible mass eigenstates. Therefore in this case we applied the
smearing of position coordinates to tame the potential at small
distance as discussed in Ref.~\cite{Isgur11}.

\subsection*{IIIc. Radiative transitions}

$E1$ radiative partial widths for meson to meson transitions were
calculated by using the following expression given in
ref.~\cite{0406228}. \small
\begin{equation}
\Gamma_{E1}(n^{2S+1}L_J\rightarrow n'^{2S'+1}L'_{J'}+\gamma)=\frac{4}{3}<e_Q>^2 \alpha \omega^3 C_{fi} \delta_{S S'} < n'^{2S'+1}L'_{J'} \mid r \mid n^{2S+1}L_J>\mid^2 \frac{E_f}{M_i}.  \label{E1}
\end{equation}
Here
\begin{equation}
<e_Q>=\frac{m_{\overline{q}}Q - m_q \overline{Q}}{m_q + m_{\overline{q}}}.
\end{equation}
$Q$($\overline{Q}$) is quark(antiquark) charge, $m_q$,
$m_{\overline{q}}$, $\alpha$, $\omega$, $E_f$, and $M_i$ represent
the quark mass, anti-quark mass, electromagnetic fine structure
constant, final photon energy, total energy of the final state
meson, and mass of initial state meson respectively, and
 \begin{equation}
C_{fi}=\max(L, L')(2 J'+1)\left \{
                           \begin{array}{ccc}
                             L' & J' & S \\
                             J & L & 1 \\
                           \end{array}
                         \right \}^2.
\end{equation}
To calculate $M1$ radiative partial widths for meson to meson transitions, the following expression~\cite{Godfrey16} was used:

\begin{multline}
\Gamma_{M1}(n^{2S+1}L_J\rightarrow n'^{2S'+1}L'_{J'}+\gamma)=\frac{\alpha}{3} \omega^3 (2J'+1)\delta_{S S'\pm 1} \frac{e_q}{m_q}<f \mid j_0(\frac{m_b}{m_q+m_b} k r)\mid i> \\ + \frac{e_b}{m_b}<f \mid j_0(\frac{m_q}{m_q+m_b} k r)\mid i>\mid^2 .
\end{multline}
\normalsize Here $j_0(x)$ is a spherical Bessel function.

\noindent In Tables (5-9) we report the calculated values of $M_1$
and $E_1$ transitions for conventional as well as hybrid $B_c$
mesons. In the $M1$ transitions the initial and final states belong
to the same orbital excitation but have different spins, and in the
$E1$ transitions the orbital quantum numbers of initial and final
states are changed but spin remains the same.

\section*{IV. Results and Conclusions}
\begin{table}\caption{Masses and radii of ground and radially excited state $B_c$ mesons. Our calculated masses are rounded to 0.001 GeV.}
\begin{center}
\begin{tabular}{|c|c|c|c|c|}
\hline
Meson & $J^{P}$& Our calculated & Exp. mass &radii\\
 & & mass & \cite{pdg17}& \\ \hline
 & &  \textrm{GeV} & \textrm{GeV} & fm\\ \hline
$B_{c} (1 ^3S_1)$ & $1^{-}$ & $6.314$ & &0.334 \\
 $B_{c}(1 ^1S_0)$ & $0^{-}$ &$6.274$ & $6.2749\pm 0.008 $ &0.318\\ \hline
 $B_{c}(2 ^3S_1)$ & $1^{-}$ & $6.855$ & &0.732\\
 $B_{c}(2 ^0S_1)$& $0^{-}$ &6.841 & $6.842 \pm 0.004 \pm 0.005$ &0.723\\ \hline
$B_{c}(3 ^3 S_1)$ & $1^{-}$ &7.206& &1.059 \\
 $B_{c}(3 ^1 S_0)$& $0^{-}$ &7.197 & &1.052\\ \hline
$B_{c}(4 ^3S_1)$ & $1^{-}$ & 7.495 & &1.342\\
 $B_{c}(4 ^1S_0)$& $0^{-}$ & 7.488 & &1.337\\ \hline

$B_{c}(1 ^3P_2) $ & $2^{+}$ &6.753 & &0.594\\
$B_{c}(1 ^{'}P_1)$ & $1^{+}$ &6.744& & \\
$B_{c}(1 P_1)$ & $1^{+}$ & 6.725 & & \\
$B_{c}(1 ^3P_0)$ & $0^{+}$ &6.701 & &0.562\\ \hline

$B_{c}(2 ^3P_2) $ & $2^{+}$ &7.111 & &0.940\\
$B_{c}(2 ^{'}P_1)$ & $1^{+}$ &7.098& & \\
$B_{c}(2 P_1)$ & $1^{+}$ & 7.105 & & \\
$B_{c}(2 ^3P_0)$ & $0^{+}$ &7.086 & &0.920\\ \hline

$B_{c}(3 ^3P_2) $ & $2^{+}$ &7.406 & &1.235\\
$B_{c}(3 ^{'}P_1)$ & $1^{+}$ &7.393& & \\
$B_{c}(3 P_1)$ & $1^{+}$ & 7.405 & & \\
$B_{c}(3 ^3P_0)$ & $0^{+}$ &7.389 & &1.220\\ \hline

$B_{c}(1 ^3D_3)$ & $3^{-}$ &6.998 & &0.793\\
$B_{c}(1 ^{'}D_2)$ & $2^{-}$ & 6.984 & & \\
$B_{c}(1 D_2)$& $2^{-}$ & 6.986 & & \\
$B_{c}(1 ^3D_1)$ & $1^{-}$ & 6.964 & &0.752\\ \hline

$B_{c}(2 ^3D_3)$ & $3^{-}$ &7.302 & &1.107\\
$B_{c}(2 ^{'}D_2)$ & $2^{-}$ & 7.293 & & \\
$B_{c}(2 D_2)$& $2^{-}$ & 7.294 & & \\
$B_{c}(2 ^3D_1)$ & $1^{-}$ & 7.280 & &1.083\\ \hline

$B_{c}(3 ^3D_3)$ & $3^{-}$ &7.570 & &1.382\\
$B_{c}(3 ^{'}D_2)$ & $2^{-}$ & 7.562 & & \\
$B_{c}(3 D_2)$& $2^{-}$ & 7.563 & & \\
$B_{c}(3 ^3D_1)$ & $1^{-}$ & 7.553 & &1.364\\ \hline

\end{tabular}
\end{center}
\end{table}

\begin{table}\caption{ Comparison of masses of ground and radially excited state $B_c$ mesons with others. Our calculated masses are rounded to 0.001 GeV.}
\begin{center}
\tabcolsep=5pt \fontsize{8}{12}\selectfont
\begin{tabular}{|c|c|c|c|c|c|c|c|c|c|}
\hline
Meson & $J^{P}$& Our & GI\cite{0406228} & EFG\cite{0210381} & \cite{1504.07538} & \cite{0606194}& \cite{lodhi}&EQ\cite{9402210}& Lattice\cite{Lattice}\\
 & & calculated & &  & & & & & \\\hline
 & &  \textrm{GeV} & \textrm{GeV} & \textrm{GeV} &\textrm{GeV}& \textrm{GeV}&\textrm{GeV} &\textrm{GeV} &\textrm{GeV}\\ \hline
$1 ^3S_1$ & $1^{-}$ &6.314 & $6.338$ & 6.332 & & 6.373 & 6.397 & 6.337 & $6.321 \pm 0.020$ \\
 $1 ^1S_0$ & $0^{-}$ &6.274 & 6.271 &6.270 & 6.277 & 6.349 & 6.356 & 6.264 &$6.280 \pm 0.030 \pm 0.190$\\ \hline

$2 ^3S_1$ & $1^{-}$ &6.855 & 6.887 & 6.881 & & 6.855 & 6.910 &6.899 & $6.990 \pm 0.080$\\
 $2 ^1S_0$ & $0^{-}$ &6.841 & 6.855 & 6.835 & 7.042 & 6.821 & 6.888 &6.856 & $6.960 \pm 0.080$\\ \hline
$3 ^3S_1$ & $1^{-}$ &7.206 & 7.272 & 7.235 & & 7.210 & &7.280 &\\
 $3 ^1S_0$ & $0^{-}$ &7.197 &7.250 & 7.193 & 7.384 & 7.175 & &7.244 &\\ \hline
$4 ^3S_1$ & $1^{-}$ &7.495 & & & & & &7.594 &\\
$4 ^1S_0$ & $0^{-}$ &7.488 & & & & & &7.562 &\\ \hline

$1 ^3P_2$ & $2^{+}$ &6.753 & 6.768 & 6.762 & &6.749 & 6.751& 6.747 &$6.783 \pm 0.03$\\
$1 ^{'}P_1$ & $1^{+}$ &6.744 & 6.750 & 6.749 & & & & 6.736& $6.765 \pm 30$\\
$1 P_1$ & $1^{+}$ & 6.725& 6.741 & 6.734 & &  & &6.730 & $6.743 \pm 30$\\
$1 ^3P_0$ & $0^{+}$ &6.701 & 6.706 & 6.699 & 6.666 & 6.715 & 6.673 & 6.7& $6.727 \pm 0.030$\\ \hline

$2 ^3P_2 $ & $2^{+}$ &7.111 & 7.164 & 7.156 & &7.153 & &7.153 &\\
$2 ^{'}P_1 $ & $1^{+}$ &7.098 & 7.15 & 7.145 & & & &7.142 &\\
$2 P_1 $ & $1^{+}$ &7.105 & 7.145 & 7.126 & & & &7.135 &\\
$2 ^3P_0$ & $0^{+}$ &7.086 & 7.122 & 7.091 & 7.207 & 7.102 & &7.108 &\\ \hline

$3 ^3P_2$ & $2^{+}$ &7.406 &  &  &  &  & & 7.153 &\\
$3 ^{'}P_1 $ & $1^{+}$ & 7.393&  &  & & & & 7.142 &\\
$3 P_1 $ & $1^{+}$ &7.405 &  &  & & & & 7.135&\\
$3 ^3P_0$ & $0^{+}$ &7.389 &  &  &  &  & & 7.108 &\\ \hline

$1 ^3D_3$ & $3^{-}$ & 6.998 & 7.045 & 7.081 & & & &7.005 & \\
$1 ^{'}D_1$ & $2^{-}$ & 6.984 & 7.036 & 7.079 & & & &7.009 &\\
$1 ^3D_1$ & $1^{-}$ & 6.964 & 7.028 & 7.072 & & & &7.012 &\\
$1 D_1$ & $2^{-}$ & 6.986 & 7.041 & 7.077 & & & & 7.012 &\\ \hline

$3 ^3D_1 $ & $1^{-}$ & 7.553 & & & & & & 7.012 &\\
$3 ^{'}D_2 $ & $2^{-}$ & 7.562 & & & & & & 7.012 &\\
$3 D_2$ & $2^{-}$ & 7.563 & & & & & & 7.009 &\\
$3 ^3D_3$ & $3^{-}$ & 7.570& & & & & & 7.005 &\\ \hline

\end{tabular}
\end{center}
\end{table}

\begin{table}\caption{Masses and radii of ground and radially excited state hybrid $B_c$ mesons. Our calculated masses are rounded to 0.001 GeV.}
\begin{center}
\begin{tabular}{|c|c|c|c|c|}
\hline
Meson & \multicolumn{2}{|c|}{$J^{P}$}& Our calculated &radii\\
 &$\varepsilon=1$& $\varepsilon=-1$  & mass & \\ \hline
 & &  & \textrm{GeV} &fm\\ \hline
$B_{c} (1 ^3S_1)$ & $1^{+}$ & $1^{-}$& $7.422$ &0.652 \\
 $B_{c}(1 ^1S_0)$ & $0^{+}$ & $0^{-}$ &$7.415$ &0.634\\ \hline
 $B_{c}(2 ^3S_1)$ & $1^{+}$ & $1^{-}$ & $7.654$ &1.017\\
 $B_{c}(2 ^0S_1)$& $0^{+}$ & $0^{-}$ &7.646 & 1.004\\ \hline
$B_{c}(3 ^3 S_1)$ & $1^{+}$  & $1^{-}$&7.874& 1.316 \\
 $B_{c}(3 ^1 S_0)$& $0^{+}$  & $0^{-}$ &7.866 &1.306\\ \hline
$B_{c}(4 ^3S_1)$ & $1^{+}$ & $1^{-}$& 8.082 &1.579\\
 $B_{c}(4 ^1S_0)$& $0^{+}$ & $0^{-}$ &8.075 &1.572\\ \hline

$B_{c}(1 ^3P_2) $ & $2^{-}$ & $2^{+}$ &7.547 &0.867\\
$B_{c}(1 ^{'}P_1)$ & $1^{-}$ & $1^{+}$ &7.547& \\
$B_{c}(1 P_1)$ & $1^{-}$ & $1^{+}$ &7.535 & \\
$B_{c}(1 ^3P_0)$ & $0^{-}$ & $0^{+}$ &7.527 &0.824\\ \hline

$B_{c}(2 ^3P_2) $ & $2^{-}$ & $2^{+}$ &7.776 &1.188\\
$B_{c}(2 ^{'}P_1)$ & $1^{-}$ & $1^{+}$&7.774& \\
$B_{c}(2 P1)$ & $1^{-}$ & $1^{+}$ &7.767 & \\
$B_{c}(2 ^3P_0)$ & $0^{-}$ & $0^{+}$ &7.759 &1.165\\ \hline

$B_{c}(3 ^3P_2) $ & $2^{-}$ & $2^{+}$ &7.990 &1.464\\
$B_{c}(1 ^{'}P_1)$ & $1^{-}$ & $1^{+}$ &7.985 & \\
$B_{c}(3 P_1)$ & $1^{-}$ & $1^{+}$ &7.985 & \\
$B_{c}(3 ^3P_0)$ & $0^{-}$ & $0^{+}$ &7.976 &1.448\\ \hline

$B_{c}(1 ^3D_3)$ & $3^{+}$ & $3^{-}$ &7.663 &1.032\\
$B_{c}(1 ^{'}D_2)$ & $2^{+}$ & $2^{-}$ &7.659 & \\
$B_{c}(1 D_2)$& $2^{+}$ & $2^{-}$ &7.660 & \\
$B_{c}(1 ^3D_1)$ & $1^{+}$ & $1^{-}$ &7.652 &0.996\\ \hline

$B_{c}(2 ^3D_3)$ & $3^{+}$ & $3^{-}$ &7.886 &1.330\\
$B_{c}(2 ^{'}D_2)$ & $2^{+}$ & $2^{-}$ &7.881 & \\
$B_{c}(2 D_2)$& $2^{+}$ & $2^{-}$ &7.882 & \\
$B_{c}(2 ^3D_1)$ & $1^{+}$ & $1^{-}$ &7.874 &1.305\\ \hline

$B_{c}(3 ^3D_3)$ & $3^{+}$ &$3^{-}$ &8.095 &1.593\\
$B_{c}(3 ^{'}D_2)$ & $2^{+}$ & $2^{-}$ &8.091 & \\
$B_{c}(3 D_2)$& $2^{+}$ & $2^{-}$ &8.091 & \\
$B_{c}(3 ^3D_1)$ & $1^{+}$ & $1^{-}$ &8.084 &1.574\\ \hline

\end{tabular}
\end{center}
\end{table}

\begin{table}\caption{The lowest masses of hybrid $B_c$ meson states with $J^P = 1^-, 1^+, 0^-, 0^+, 2^-, 2^+$.}
\begin{center}
\begin{tabular}{|c|c|c|}
\hline
$J^P$ & Our calculated mass & QCD Sum rule\cite{1306.3486} \\  \hline
 & \textrm{GeV} & \textrm{GeV} \\  \hline
$1^-$ &\multirow{2}{*}{7.422}& $6.83 \pm 0.08 \pm 0.01 \pm 0.07$ \\
$1^+$ &  & $7.77 \pm 0.06 \pm 0.05 \pm 0.13$ \\ \hline
$0^-$ & \multirow{2}{*}{7.415} & $6.90 \pm 0.12 \pm 0.01 \pm 0.09$ \\
$0^+$ &  & $7.37 \pm 0.12 \pm 0.07 \pm 0.12$ \\ \hline
$2^-$ &\multirow{2}{*}{7.547} & $7.15 \pm 0.08 \pm 0.05 \pm 0.09$ \\
$2^+$ &  & $7.67 \pm 0.07 \pm 0.02 \pm 0.09$ \\ \hline
\end{tabular}
\end{center}
\end{table}

The aim of the present paper is to study conventional and hybrid
$B_c$ mesons. For this purpose, we calculate the masses, radii and
radiative transitions for ground and radially excited conventional
and hybrid $B_c$ meson states. In Table 1 and Table 3, our
calculated masses and radii are reported for the ground and radially
excited states of conventional and hybrid $B_c$ mesons respectively.
 Only two $B_c$ meson states ($B_c(1S), B_c(2S)$) are known experimentally.
The experimental masses of these states are given in 4th column of
Table 1. Table 1 and 3 show that the mass and radii of the
conventional and hybrid $B_c$ mesons monotonically increase with
radial and orbital excitations. The similar results are obtained for
the charmonium and bottomonium mesons in refs.
\cite{Nosheen11,Nosheen14,Nosheen16}. In Table 2, we compare our
calculated masses of conventional $B_c$ mesons with others
\cite{9402210,0210381,0406228,1504.07538,Lattice,0606194,lodhi}. It
is observed our results well agree with the calculated spectrum by
others as mentioned in Table 2.
 In Table 3, the calculated masses of hybrid $B_c$ mesons are
reported for the same values of $n$, $L$, and $S$ as used for the
conventional mesons. In order to distinguish hybrids from
non-hybrids, we use here a workable notation of adding
 a superscript
$h$ to the symbol of the conventional meson with the same $n$, $L$,
and $S$. The same notation is already used in \cite{Nosheen16}.
These results show that for the same quantum numbers ($n$, $L$, and
$S$) the mass of a hybrid meson is significantly greater than that
of the corresponding conventional meson. It is noted that $J^{P}$ of
each hybrid meson is also different from the corresponding
conventional meson for same $L$ and $S$. This difference arises
because of the angular momentum of the gluonic field which
contributes in the hybrid case. It is also noted that the gluonic
potential $\Pi _{u}$ applied in this work allows two possible value
of $\epsilon $ in eq. (7). As a result we obtain two degenerate
hybrid states with opposite values of parity. Observing the results
reported in Table 3, it is found that the lightest hybrid $B_c$
meson state has mass 7.422 GeV with $J^{P}=1^{+}(1^{-})$ which is
greater than the lowest conventional $B_c$ meson state. In Ref.
\cite{1306.3486} the masses of $B_c$ hybrid mesons having $J^P =
0^-, 0^+, 1^-, 1^+, 2^-, 2^+$ are calculated using the QCD sum rule.
The comparison of our results with that of Ref.~\cite{1306.3486} is
provided in Table 4. This Ref. predicts that the lightest $B_c$
hybrid state is $1^-$ with a mass of $6.83 \pm 0.08 \pm 0.07$ GeV,
whereas our potential model predicts its mass to be $7.422$ GeV.

\noindent In Tables (5-9), our calculated electric dipole (E1) and
magnetic dipole (M1) transitions are reported. In  4th column of
tables (5-8) E1 radiative transitions for conventional to
conventional $B_c$ mesons are reported, whereas hybrid to hybrid
radiative transitions are reported in  5th column of these tables.
M1 transitions from conventional to conventional and hybrid to
hybrid $B_c$ meson are reported in  4th and 5th column of Table 9.
It is noted that the E1 radiative transitions are typically of order
of 1 to 100 keV except for the $3P \rightarrow 1S$ transitions,
whereas the M1 transitions are reduced due to the presence of mass
factor in the denominator of the formula. Nevertheless M1
transitions have been useful in observing spin singlet states that
are difficult to observe otherwise. We observe that the radiative
transition rates from conventional to conventional mesons are higher
than those for the hybrid to hybrid transitions with the same
quantum numbers of the initial and final states, except few
transitions ($2 ^3 P_2\rightarrow 1 ^3 D_3$, $3 ^3P_2\rightarrow 2
^3 D_3$,  $3 ^3P_2\rightarrow 2 ^3 D_1$). Generally both E1 and M1
transition rates are also very small when the transitions occur
between the states with close masses because of the reduced value of
$E_{\gamma}$. We find same behavior in the case of radiative
transitions of hybrid $b\bar{c}$ states. To our knowledge, hybrid
$B_c$ mesons masses are studied only using the QCD sum rules in Ref.
\cite{1306.3486}. In this Ref. masses are predicted only for six
hybrid states ($0^+, 0^-, 1^+, 1^-,2^+,2^-$), whereas we provide
masses of complete spectrum with several radial and orbital
excitations.

This work can be helpful in $B_c$ meson searches at laboratories like BESIII, PANDA and LHCb.
\begin{table}\caption{$S\rightarrow P$ E1 radiative transitions. The masses are taken from above mentioned Table 1 and 3; we use the experimental masses if known. Otherwise, theoretically calculated masses are used.}
\tabcolsep=4pt
\fontsize{9}{11}\selectfont
\begin{center}
\begin{tabular}{|c|c|c|c|c|}
\hline
Transition & Initial & Final & Our calculated & Our calculated \\
 & Meson & Meson & $\Gamma_{E1}$ & $\Gamma_{E1}$ for hybrids \\
  & & & keV & keV \\ \hline
$2S\rightarrow 1P$ & $2 ^3S_1$ & $1^3P_2$ & 2.092 & 2.384 \\
 &  & $1^3P_1^{'}$ & 3.15 &1.45  \\
 &  & $1^3P_1$ & 2.52 &0.999  \\
&  & $1^3P_0$ & 1.395 & 0.804 \\
&$2 ^1S_0$ & $1^1P_1^{'}$ &3.44 & 1.93\\
&$2 ^1S_0$ & $1^1P_1$ &11.59 & 5.39\\ \hline
$3S\rightarrow 2P$ & $ 3 ^3S_1$ & $ 2^3P_2$ & 1.713  & 1.858 \\
  &  & $2^3P_0$ & 0.672 & 0.594 \\
$3S\rightarrow 1P$ & $ 3^3S_1$ & $1^3P_2$ &161.840  & 63.918 \\
 & & $1^3P^{'}_1$ &0.104 & 0.004 \\
 &  & $1^3P_1$ &0.058 & 0.002 \\
 &  & $1^3P_0$ &43.877 & 15.257 \\
 & $ 3^1S_0$ & $1^1P^{'}_1$ &0.624 & 0.060 \\
 & & $1^1P_1$ & 1.399 & 0.132 \\ \hline
$4S\rightarrow 3P$ & $4 ^3S_1$ & $3^3P_2$ & 1.412 & 1.549 \\
 &  & $3^3P_0$ & 0.469 & 0.470 \\
$4S\rightarrow 2P$ & $4^3S_1$ & $2^3P_2$ & 101.987 & 53.254 \\
  &  & $2^3P_0$ & 24.319 & 12.436 \\
$4S\rightarrow 1P$ & $4^3S_1$ & $1^3P_2$ & 648.85 & 263.851 \\
 &  & $1^3P_0$ & 155.86 & 58.720 \\ \hline
\end{tabular}
\end{center}
\end{table}
\begin{table}\caption{1P and 2P E1 radiative transitions.}
\tabcolsep=4pt
\fontsize{9}{11}\selectfont
\begin{center}
\begin{tabular}{|c|c|c|c|c|}
\hline
Transition & Initial & Final & Our calculated & Our calculated \\
 & Meson & Meson & $\Gamma_{E1}$ & $\Gamma_{E1}$ for hybrids \\
  & & & keV & keV \\ \hline
$1P\rightarrow 1S$ & $1^3P_2$ & $1^3S_1$ &87.562 & 2.317\\
 &  $1^3 P^{'}_1$ & & 73.71  & 3.10 \\
 &  $1 ^3 P_1$ & & 72.48  & 1.23 \\
 &  $1^3P_0$ & & 61.347 & 1.362\\
  &  $1^1 P^{'}_1$ & $1^1S_0$ & 41.82  &1.84  \\
 &$1 ^1 P_1$ & & 74.17 & 2.76 \\ \hline
$2P\rightarrow 2S$ & $2^3P_2$ & $2^3S_1$ & 18.660 & 2.132\\
 & $2^3 P^{'}_1$ & & 40.35 & 12.70\\
& $2^3 P_1$ & & 21.98 &5.33 \\
 & $2^3P_0$ & & 13.936 & 1.374\\
&$2^1P^{'}_1$ & $2^1S_0$ &21.40 & 7.09\\
&$2^1 P_1$ & $2^1S_0$ &46.16 &11.97 \\ \hline
$2P\rightarrow 1S$ & $2 ^3P_2$ & $1^3S_1$ & 464.552 & 48.376\\
 & $2 ^3 P^{'}_1$ & & 17.99 & $1.2 \times 10^{-10}$\\
 & $2 ^3 P_1$ & & 9.24 & $5.3 \times 10^{-11}$\\
 & $2^3P_0$ & &426.574 & 42.028\\
&$2^1 P^{'}_1$ & $1^1S_0$ &12.74 &0.091  \\
&$2^1 P_1$ & $1^1S_0$ &25.93 & 0.18 \\\hline
$2P\rightarrow 1D$ & $2^3P_2$ & $1^3D_3$ & 2.821 & 2.841 \\
 &  & $1^3 D^{'}_2$ & 0.91 & 0.81 \\
 &  & $1^3 D_2$ & 1.30 & 1.18 \\
 & $2 ^3 P^{'}_1$ & $1^3 D^{'}_2$ & 2.36 & 2.82\\
&  & $1^3D_1$ & 2.74 & 2.45\\
 & $2^1P^{'}_1$ & $1^1 D_2$ & 2.31 & 1.83\\
 &$2^3 P_1$ & $1^3 D^{'}_2$ &1.41 & 1.17\\
&  & $1^3D_1$ & 1.60 & 1.03\\
&$2 ^1 P_1$ & $1 ^1 D_2$ &3.66 & 3.03\\
  &  $2^3P_0$ & $1^3D_1$  &4.276 & 2.862\\ \hline
\end{tabular}
\end{center}
\end{table}

\begin{table}\caption{3P E1 radiative transitions}
\tabcolsep=4pt
\fontsize{9}{11}\selectfont
\begin{center}
\begin{tabular}{|c|c|c|c|c|}
\hline
Transition & Initial & Final & Our calculated & Our calculated \\
 & Meson & Meson & $\Gamma_{E1}$ & $\Gamma_{E1}$ for hybrids \\
  & & & keV & keV \\ \hline
$3P\rightarrow 3S$ & $3 ^3P_2$ & $3^3S_1$ & 9.135 & 1.850 \\
 & $3 ^3P^{'}_1$ & & 34.375 & 14.759 \\
 & $3 ^3P_1$ & & 20.671 &  \\
 & $3^3P_0$ & &7.076 & 1.268\\
 & $3 ^1P^{'}_1$ & $3^1 S_0$ & 18.367 & 8.362 \\
 & $3 ^1 P_1$ &  & 43.613 & 16.629 \\ \hline
$3P\rightarrow 2S$ & $3 ^3P_2$ & $2^3S_1$ &169.486 & 41.775 \\
 & $3^3P_0$ &  &155.404 & 36.958\\
$3P\rightarrow 1S$ & $3 ^3P_2$ & $1^3S_1$ &1091.840 & 187.423\\
 & $3^3P_0$ &  &1047.780 & 174.716 \\ \hline
$3P\rightarrow 2D$ & $3 ^3P_2$ & $2^3D_3$ &2.198 &2.236\\
 &  &$2^3 D^{'}_2$ &1.469 & 1.418 \\
&  &$2^3 D_2$ &2.151 &2.075 \\
 &  &$2^3D_1$ &0.046 & 0.073 \\
 & $3^3 P_0$ &$2^3D_1$ &3.050 & 2.516\\ \hline
$3P\rightarrow 1D$ & $3^3P_2$ & $1^3D_3$ &121.442 & 64.782\\
 &  &$1^3D_2$ &23.940 & 11.953 \\
 &  &$1^3D_1$ &1.816 & 0.846\\
& $3^3P_1$ & $1^3D_2$ &115.076 & 57.294\\
 &  &$1^3D_1$ &43.728 & 20.294 \\
 & $3^3 P_0$ &$1^3D_1$ &162.715 & 74.888 \\
 &$3 ^1P_1$ & $1^1D_2$ &148.539 & 76.182 \\ \hline
\end{tabular}
\end{center}
\end{table}

\begin{table}\caption{1D and 2D E1 radiative transitions.}
\tabcolsep=4pt
\fontsize{9}{11}\selectfont
\begin{center}
\begin{tabular}{|c|c|c|c|c|}
\hline
Transition & Initial & Final & Our calculated & Our calculated \\
 & Meson & Meson & & $\Gamma_{E1}$ for hybrids \\
  & & & keV & keV \\ \hline
$1D\rightarrow 1P$ & $1 ^3D_3$ & $1^3P_2$ & 19.790 & 2.192 \\
  & $1 ^3D_2$ & $1 ^3 P^{'}_1$ & 24.16 & 5.63 \\
 &  & $1 ^3 P_1$ & 15.09 & 3.80\\
&$1^3 D_1$ & $1^3 P_1$ &10.90 & 2.89 \\
 &  & $1 ^3 P^{'}_1$ & 17.08 & 4.18\\
 & $1^1 D^{'}_2$ & $1^1 P_1$ &39.19 & 6.65\\
 &$1^1 D^{'}_2$ & $1 P^{'}_1$& 15.74  & 2.47\\
& & $1^3 P_0$ &13.426 & 1.555\\
 \hline
$2D\rightarrow 2P$ & $2 ^3D_3$ & $2^3P_2$ &9.706 & 1.896\\
 & $2 ^3D_2$ & $2^3P_2$ & 2.091 & 0.418\\
 &  & $2^3P_1$ & 7.206 & 1.469 \\
&$2^3D_1$ & $2^3 P_2$ &0.188 & 0.037\\
& & $2^3 P_1$ &3.273 & 0.667\\
& & $2^3 P_0$ &5.575 & 1.197\\
&$2^1 D_2$ & $2^1P_1$ &9.949 & 1.943\\ \hline

$2D\rightarrow 1P$ & $2 ^3D_3$ & $1^3P_2$ & 201.438 & 51.191\\
 & $2 ^3D_2$ & $1^3P_2$ &47.949 & 12.300 \\
 &  & $1^3P_1$ &158.935 & 39.142 \\
&$2^3D_1$ & $1^3 P_2$ &4.982 & 1.281\\
& & $1^3 P_1$ & 82.772 & 20.417 \\
& & $1^3 P_0$ &129.345 & 30.582 \\
&$2^1 D_2$ & $1^1P_1$& 209.866 & 51.634\\ \hline

\end{tabular}
\end{center}
\end{table}

\begin{table}\caption{ M1 radiative transitions.}
\tabcolsep=4pt
\fontsize{9}{11}\selectfont
\begin{center}
\begin{tabular}{|c|c|c|c|c|}
\hline
Transition & Initial & Final & Our calculated & Our calculated \\
 & Meson & Meson &$\Gamma_{M1}$ & $\Gamma_{M1}$ for hybrids\\
   & & & keV & keV \\ \hline
$1S$ & $1^3S_1$ & $1^1S_0$ & 0.027 & $1.6\times 10^{-4}$ \\ \hline
2S &$2^3 S_1$ & $2^1S_0$ & $1.6\times 10^{-6}$ & $2.5\times 10^{-6}$ \\
 &  & $1^1S_0$ & 0.367 & 0.015\\
&$2 ^1S_0$ & $1^3S_1$ &0.006 & 0.003 \\ \hline
3S & $3 ^3S_1$ & $3^1S_0$ & 0.00032 & $2.1 \times 10^{-4}$ \\
 &  & $2^1S_0$ &0.096 & 0.023\\
 &  & $1^1S_0$ & 0.431 & 0.016\\
&$3 ^1S_0$ & $2^3S_1$ &0.0046 & $ 2.6\times 10^{-4}$ \\
 &  & $1^3S_1$ &0.000646 & 0.020 \\ \hline
\end{tabular}
\end{center}
\end{table}

\begin{table} \caption{ Comparison of our calculated $E_1$ transitions with others.}
\tabcolsep=4pt
\fontsize{9}{11}\selectfont

\begin{center}
\begin{tabular}{|c|c|c|c|c|c|}
\hline
Transition & Initial & Final & Our calculated & $\Gamma_{E1}$ & $\Gamma_{E1}$ \\
 & Meson & Meson & $\Gamma_{E1}$ & \cite{0406228}  & \cite{0210381} \\
  & & & keV & keV & keV \\ \hline
$2S\rightarrow 1P$ & $2^3S_1$ & $1^3P_2$ & 2.092 &5.7 & 7.59 \\
 &  & $1^3P_0$ & 1.395 &2.9  &5.53 \\
  &  & $1^3P_1^{'}$ & 3.15 & 0.7& 0.74\\
 &  & $1^3P_1$ & 2.52 &4.7 & 7.65 \\
&$2 ^1S_0$ & $1^1P_1^{'}$ &3.44 & 6.1 & 4.40\\
&$2 ^1S_0$ & $1^1P_1$ &11.59 & 1.3 & 1.05\\ \hline
$1P\rightarrow 1S$ & $1 ^3P_2$ & $1^3S_1$ &87.562 & 83 & 122 \\
 &  $1^3 P^{'}_1$ & & 73.71  & 11 & 13.7\\
 &  $1 ^3 P_1$ & & 72.48  & 60& 87.1 \\
 &  $1^3P_0$ & & 61.347 & 55&\\
  &  $1^1 P^{'}_1$ & $1^1S_0$ & 41.82  & 80 & 147 \\
 &$1 ^1 P_1$ & & 74.17 &13& 18.4 \\ \hline
$2P\rightarrow 2S$ & $2 ^3P_2$ & $2^3S_1$ & 18.660  & 55 &75.3 \\
 & $2^3 P^{'}_1$ & & 40.35 & 5.5& 1.49\\
& $2^3 P_1$ & & 21.98 & 45& 45.3 \\
 & $2^3P_0$ & & 13.936 & 42& 34\\
&$2^1P^{'}_1$ & $2^1S_0$ &21.40 &52 &90.5\\
&$2^1 P_1$ & $2^1S_0$ &46.16 &5.7 & 13.8\\ \hline
$2P\rightarrow 1S$ & $2 ^3P_2$ & $1^3S_1$ & 464.552 &14 & \\
  & $2 ^3 P^{'}_1$ & & 17.99 &0.6 & \\
 & $2 ^3 P_1$ & & 9.24 & 5.4&\\
 & $2^3P_0$ & &426.574 &1.0 &\\
&$2^1 P^{'}_1$ & $1^1S_0$ &12.74 & 19&  \\
&$2^1 P_1$ & $1^1S_0$ &25.93 &2.1 & \\\hline
$2P\rightarrow 1D$ & $2 ^3P_2$ & $1^3D_3$ & 2.821 &6.8 & 2.08 \\
 &  & $1^3 D^{'}_2$ & 0.91 &0.7 &0.139 \\
 &  & $1^3 D_2$ & 1.30 &0.6 &0.285 \\
 & $2 ^3 P^{'}_1$ & $1^3 D^{'}_2$ & 2.36 & 5.5&10.4\\
&  & $1^3D_1$ & 2.74 &0.2 & 0.070\\
 & $2^1P^{'}_1$ & $1^1 D_2$ & 2.31 &1.3 & 0.172\\
 &$2^3 P_1$ & $1^3 D^{'}_2$ &1.41 & 0.8& 0.023\\
&  & $1^3D_1$ & 1.60 &1.6 & 0.204\\
&$2 ^1 P_1$ & $1 ^1 D_2$ &3.66 & 3.6& 0.517\\
&  $2^3P_0$ & $1^3D_1$  &4.276 & 4.2 & 0.041 \\ \hline
 $1D\rightarrow 1P$ & $1^3D_3$ & $1^3P_2$ & 19.790 &78 &149 \\
 & $1 ^3D_2$ & $1 ^3 P^{'}_1$ & 24.16 &15 & 14.9 \\
 &  & $1 ^3 P_1$ & 15.09 & 64& 139\\
&$1^3 D_1$ & $1^3 P_1$ &10.90 & 28& 65.3\\
 &  & $1 ^3 P^{'}_1$ & 17.08 & 4.4& 7.81\\
 & $1^1 D^{'}_2$ & $1^1 P_1$ &39.19 &7 & 7.1\\
 &$1^1 D^{'}_2$ & $1 P^{'}_1$& 15.74  &63 & 143\\
& & $1^3 P_0$ &13.426 & 55 & 133 \\ \hline
\end{tabular}
\end{center}
\end{table}

\begin{table} \caption{ Comparison of our calculated $M_1$ transitions with others.}
\tabcolsep=4pt
\fontsize{9}{11}\selectfont

\begin{center}
\begin{tabular}{|c|c|c|c|c|c|}
\hline
Transition & Initial & Final & Our calculated & $\Gamma_{M1}$ & $\Gamma_{M1}$ \\
 & Meson & Meson & $\Gamma_{M1}$ & \cite{0406228}  & \cite{0210381} \\
  & & & keV & keV & keV \\ \hline

$1S$ & $1^3S_1$ & $1^1S_0$ & 0.027 &0.08 & 0.073 \\ \hline
2S &$2^3 S_1$ & $2^1S_0$ & $1.6\times 10^{-6}$ & 0.01& 0.03 \\
 &  & $1^1S_0$ & 0.367 &0.6 &0.141\\
&$2 ^1S_0$ & $1^3S_1$ &0.006 & 0.3 & 0.160 \\ \hline
3S & $3^3S_1$ & $3^1S_0$ & 0.00032 &0.003 & \\
 &  & $2^1S_0$ &0.096 &0.2 & \\
 &  & $1^1S_0$ & 0.431 &0.6 & \\
&$3 ^1S_0$ & $2^3S_1$ &0.0046 &0.06 & \\
 &  & $1^3S_1$ &0.000646 &4.2 &\\ \hline
\end{tabular}
\end{center}
\end{table}

\end{document}